\documentclass[a4paper,11pt]{article}

\pdfoutput=1 
\usepackage{amsmath}
\usepackage{graphicx}

\usepackage[english]{babel}
\usepackage[utf8]{inputenc}
\usepackage{pdflscape}
\usepackage{enumerate}
\usepackage{amsbsy}
\usepackage{amsmath} 
\usepackage{graphics}
\usepackage{mathrsfs}
\usepackage{wrapfig}
\usepackage{mathtools}

\usepackage{amsfonts}
\usepackage{pstricks}
\usepackage{color}
\usepackage{setspace}

\usepackage{jcappub_ver} 

\newcommand{\bea}{\begin{eqnarray}} \newcommand{\eea}{\end{eqnarray}}
\newcommand{\el}{\nonumber \\}
\newcommand{\re}[1]{(\ref{#1})}

\newcommand{\pat}{\partial}

\renewcommand{\a}{\alpha}
\renewcommand{\b}{\beta}
\renewcommand{\c}{\gamma}

\newcommand{\ha}{\frac{1}{2}}

\newcommand{\rmd}{\mathrm{d}}

\newcommand{\ie}{i.e.\ }

\newcommand{\Ueff}{U}
\newcommand{\Ub}{\bar{U}}
\newcommand{\epsb}{\bar{\epsilon}}
\newcommand{\etab}{\bar{\eta}}
\newcommand{\zetab}{\bar{\zeta}}

\title{Inflation with $R^2$ term in the Palatini formulation}

\author[a]{Vera-Maria~Enckell,}
\author[a]{Kari~Enqvist,}
\author[a,b]{Syksy~R{\"a}s{\"a}nen}
\author[a]{and Lumi-Pyry~Wahlman}

\affiliation[a]{University of Helsinki, Department of Physics and Helsinki Institute of Physics \\ P.O. Box 64, FIN-00014 University of Helsinki, Finland}

\affiliation[b]{Birzeit University, Department of Physics \\
P.O. Box 14, Birzeit, West Bank, Palestine}

\emailAdd{vera-maria.enckell@helsinki.fi}
\emailAdd{kari.enqvist@helsinki.fi}
\emailAdd{syksy.rasanen@iki.fi}
\emailAdd{pyry.wahlman@helsinki.fi}

\abstract{
We study scalar field inflation in $F(R)$ gravity in the Palatini formulation of general relativity. Unlike in the metric formulation, in the Palatini formulation $F(R)$ gravity does not introduce new degrees of freedom. However, it changes the relations between existing degrees of freedom, including the inflaton and spacetime curvature. Considering the case $F(R)=R+\a R^2$, we find that the $R^2$ term decreases the height of the effective inflaton potential. By adjusting the value of $\a$, this mechanism can be used to suppress the tensor-to-scalar ratio $r$ without limit in any scalar field model of inflation without affecting the spectrum of scalar perturbations.}

\begin{document}

\begin{flushleft}
	\hfill		 HIP-2018-19/TH \\
\end{flushleft}

\maketitle
  
\setcounter{tocdepth}{2}

\setcounter{secnumdepth}{3}

\section{Introduction} \label{sec:intro}

A model of inflation is usually defined by the choice of the scalar fields involved, their kinetic terms and potentials, and couplings to gravity and other fields. However, we also have to specify the form of the gravitational action and choose the gravitational degrees of freedom. One example of the latter is the choice between the metric and the Palatini formulation.

In the usual formulation of general relativity, the connection is defined in terms of the metric. In the Palatini formulation, the metric and the connection are instead taken to be independent degrees of freedom \cite{einstein1925, ferraris1982}. In the case of the Einstein--Hilbert action plus matter coupled to the metric, but not its derivatives nor the connection, the equation of motion for a symmetric connection yields the Levi--Civita connection. The metric formulation and the Palatini formulation are then physically equivalent. When matter couples to the derivatives of the metric or to the connection, this is no longer true. In particular, this is the case when a scalar field is coupled directly to the Ricci scalar \cite{Lindstrom:1976a, Lindstrom:1976b, Bergh:1981, Bauer:2008zj, Bauer:2010jg, Koivisto:2005, Rasanen:2017, Enckell:2018, Markkanen:2017, Kozak:2018, Jarv:2017}, as is the case in Higgs inflation \cite{Bezrukov:2007ep, Bauer:2008zj, Bauer:2010jg, Rasanen:2017, Enckell:2018, Markkanen:2017}. The two formulations also differ, when the gravitational action is more complicated than the Einstein--Hilbert action \cite{Buchdahl:1960, Buchdahl:1970, ShahidSaless:1987, Flanagan:2003a, Flanagan:2003b, Sotiriou:2006, Sotiriou:2008, Olmo:2011, Borunda:2008, Querella:1999, Cotsakis:1997, Jarv:2018, Conroy:2017, Li:2007, Li:2008, Exirifard:2007}. 

Perhaps the simplest extended gravitational action is given by replacing the Einstein--Hilbert action by a function $F(R)$ of the Ricci scalar. Higher order terms in the Riemann tensor are generated by quantum corrections and should thus in general be included. In the metric formulation, the $F(R)$ theory is special in that it does not suffer from the Ostrogradski instability \cite{Simon:1990ic, MuellerHoissen:1991, Woodard:2006nt}. A non-linear function $F(R)$ is equivalent to having an extra scalar field in addition to normal gravity \cite{Sotiriou:2008}. If we restrict to terms that are at most dimension four, this reduces to the Einstein--Hilbert action plus an $R^2$ term. This term (coming from the trace anomaly) was used in the first model of inflation \cite{Starobinsky:1980te}. If a scalar matter field is also present, there is the possibility of two-field inflation. Inflation with the $R^2$ term and the Higgs field has been studied in the metric formulation in \cite{Barbon:2015, Wang:2017fuy, Ema:2017rqn, He:2018gyf, Ghilencea:2018rqg, Salvio:2015kka, Salvio:2017oyf, Calmet:2016fsr, Gorbunov:2018llf}.

In contrast, in the Palatini formulation a non-linear $F(R)$ does not introduce new degrees of freedom \cite{ShahidSaless:1987, Sotiriou:2006, Sotiriou:2008, Olmo:2011}. However, it does change the relation between matter and curvature, and can thus modify inflationary predictions.

We study inflation with a scalar field in the Palatini formulation when the gravitational action contains an $R^2$ term. In section \ref{sec:mec} we begin from an action with an $F(R)$ term, considering both minimal and non-minimal coupling to gravity. After field redefinitions and a conformal transformation we obtain an action for a minimally coupled and canonically normalised scalar field in the Einstein frame. We derive the effective potential and calculate the slow-roll parameters and observables, and compare to the case without the $R^2$ term. We show that the spectral index and the amplitude of the scalar perturbations are unaffected, but the tensor-to-scalar ratio is suppressed, regardless of whether the inflaton is minimally or non-minimally coupled. In section \ref{sec:conc} we present our conclusions.

\section{Inflation with an \texorpdfstring{$R^2$}{R\textasciicircum 2} term} \label{sec:mec}

\subsection{The action and the potential}

Let us consider a scalar field $h$, the metric $g_{\a\b}$ and a symmetric connection $\Gamma^{\c}_{\a\b}$, treated as independent variables. We thus consider an action of the form
\bea \label{action1}
  S &=& \int\rmd^4 x \sqrt{-g} \left[ \ha F(R) + \ha G(h) R - \ha g^{\a\b} \pat_\a h \pat_\b h - V(h) \right]
\eea
where $g$ is the determinant of $g_{\a\b}$ and $R=g^{\a\b} R^{\c}_{\ \, \a\c\b}(\Gamma, \pat\Gamma)$ is the Ricci scalar (the Riemann tensor is built from the connection alone). We can rewrite the gravitational part of \eqref{action1} in terms of an action linear in $R$ plus a new scalar field as
\bea \label{action2}
  S &=& \int\rmd^4 x \sqrt{-g} \left[ \ha G(h) R + \ha \{ F(\phi) + F'(\phi) (R - \phi) \} - \ha g^{\a\b} \pat_\a h \pat_\b h - V(h) \right] \ ,
\eea
where $\phi$ is an auxiliary field \cite{Sotiriou:2008}. Variation with respect to $\phi$ yields the constraint $\phi=R$ (if $F''\neq0$), and substituting this result back to \re{action2} we recover the original action \re{action1}. Redefining the auxiliary field as $\varphi \equiv F'(\phi)$ we obtain the action
\bea \label{action3}
  S &=& \int\rmd^4 x \sqrt{-g} \left[ \ha [ \varphi + G(h) ] R - W(\varphi) - \ha g^{\a\b} \pat_\a h \pat_\b h - V(h) \right] \ ,
\eea
where the potential for $\varphi$ is
\bea \label{W}
  W(\varphi) \equiv \ha \{ \phi(\varphi) \varphi - F[\phi(\varphi)] \} \ .
\eea
As there is no kinetic term for $\varphi$, its equation of motion reduces to a constraint, and the action describes a single scalar field $h$ non-minimally coupled to gravity. As usual, we make a conformal transformation to obtain a minimally coupled scalar field. The transformation depends on both $h$ and $\varphi$
\bea
  g_{\a\b} \rightarrow \Omega^2 g_{\a\b} = [ \varphi + G(h) ] g_{\a\b} \ .
\eea
The action becomes
\bea \label{action4}
  S &=& \int\rmd^4 x \sqrt{-g} \left[ \ha R - \ha \frac{1}{\varphi + G(h)} g^{\a\b} \pat_\a h \pat_\b h - \hat V(h, \varphi) \right] \ ,
\eea
where we have chosen units such that the Planck mass is unity, and the conformally transformed potential is
\bea \label{U}
  \hat V(h, \varphi) \equiv \frac{V(h) + W(\varphi)}{[\varphi + G(h)]^2} \ .
\eea

Obtaining a canonical kinetic term for the scalar field is a bit involved, because the prefactor of the kinetic term depends on $\varphi$. Let us specialise to the case
\bea \label{FR}
  F(R) = R + \alpha R^2 \ ,
\eea
where $\alpha$ is a constant (\ie we consider only terms up to dimension 4). The potential then becomes
\bea \label{U2}
  \hat V(h, \varphi) = \frac{1}{[\varphi + G(h)]^2} \left[ V(h) + \frac{(\varphi-1)^2}{8 \a} \right] \ .
\eea

Varying \re{action4} with respect to $\varphi$, we get a constraint equation with the solution
\bea \label{varphi}
  \varphi = \frac{ 1 + G(h) + 8 \a V(h) + 2 \a G(h) \pat^\a h \pat_\a h }{ 1 + G(h) - 2 \a \pat^\a h \pat_\a h } \ .
\eea
In the limit of small $\a$, we get $\varphi\simeq1$, and \re{U2} reduces to the usual potential for $h$. In general, we can insert \re{varphi} into \re{U2} to eliminate $\varphi$ and write the potential in terms of $h$ and $\pat^\a h \pat_\a h$. Inserting the potential back into the action \re{action4}, we obtain
\bea \label{action5}
  S &=& \int\rmd^4 x \sqrt{-g} \left[ \ha R - \ha \frac{1}{ (1+G) (1+8\a \Ub) } \pat^\a h \pat_\a h \right. \el
 && \left. + \frac{\a}{2} \frac{1}{(1+G)^2 (1+8\a \Ub)} ( \pat^\a h \pat_\a h )^2 - \frac{\Ub}{1+8\a \Ub} \right] \ ,
\eea
where we have defined
\bea \label{Ub}
  \Ub(h) \equiv \frac{V(h)}{[1+G(h)]^2} \ .
\eea
Note that $\Ub$ is the usual Einstein frame potential in the case where we do not have the $R^2$ term \cite{Bezrukov:2007ep, Bauer:2008zj}. We can now redefine the field through the equation
\bea \label{ft}
  \frac{\rmd h}{\rmd\chi} = \pm \sqrt{ ( 1 + G ) ( 1 + 8 \a \Ub ) } \ ,
\eea
to obtain a canonical kinetic term for $\chi$ (apart from the $( \pat^\a \chi \pat_\a \chi )^2$ contribution). In terms of the new field $\chi$ the action reads
\bea \label{action6}
  S &=& \int\rmd^4 x \sqrt{-g} \left[ \ha R - \ha \pat^\a \chi \pat_\a \chi + \frac{\a}{2} (1+8\a \Ub) ( \pat^\a \chi \pat_\a \chi )^2 - \frac{\Ub}{1+8\a \Ub} \right] \ .
\eea
The $R^2$ term has been translated into a higher order kinetic term and a modification of the potential. Because a negative $\a$ would lead to negative kinetic energy, implying an unstable system, we take $\a>0$.

For small $\a$, the effect of the $\a R^2$ term reduces to the small correction
$\a [ \frac{1}{2} ( \pat^\a \chi \pat_\a \chi )^2 + 8 \Ub^2]$, which is what we could have naively expected. In general the effective potential reads
\bea \label{Ueff}
  \Ueff \equiv \frac{\Ub}{1 + 8 \a \Ub} \ .
\eea
Regardless of the shape of $\Ub$, the $R^2$ term decreases the height of the effective potential. Even if $\Ub$ grows without limit, the effective potential becomes flat approaching the value $1/8\a$. Thus the $R^2$ term can convert any potential $\Ub(h)$ into a plateau or a hilltop potential. However, when calculating the observables we have to take into account that the rate of change of the field is also modified, as \re{ft} shows. Let us now write the equations of motion and consider the slow-roll limit to see what happens in detail.

\subsection{Equations of motion and observables}

The Einstein equation for the action \re{action6} reads
\bea \label{Einstein}
  G_{\a\b} &=& \pat_\a \chi \pat_\b \chi [ 1 - 2 \a (1+8\a \Ub) \pat^\c \chi \pat_\c \chi ] \el
  && + g_{\a\b} \left[ - \frac{1}{2} \pat^\c \chi \pat_\c \chi + \frac{\a}{2} (1+8\a \Ub) ( \pat^\c \chi \pat_\c \chi)^2 - \Ueff \right] \ ,
\eea
where $G_{\a\b}$ is the Einstein tensor. The equation of motion for $\chi$ is
\bea \label{eomchi}
  0 &=& [ 1 - 2 \a (1+8\a \Ub) \pat^\a \chi \pat_\a \chi ] \Box \chi - 4 \a (1+8\a \Ub) \nabla_\a \nabla_\b \chi \pat^\a \chi \pat^\b \chi \el
  && -12 \a^2 ( \pat^\a \chi \pat_\a \chi )^2 \Ub' - \Ueff' \ ,
\eea
where prime denotes derivative with respect to $\chi$.

In the Friedmann--Robertson--Walker case (with zero spatial curvature) these equations reduce to
\bea \label{FRW}
  3 H^2 &=& \frac{1}{2} [ 1 + 3 \a ( 1+8\a \Ub ) \dot\chi^2 ] \dot\chi^2 + \Ueff \el  
  0 &=& [ 1+ 6 \a ( 1 + 8 \a \Ub ) \dot\chi^2 ] \ddot\chi + 3 [ 1 + 2 \a ( 1 + 8 \a \Ub ) \dot\chi^2 ] H \dot\chi + 12 \a^2 \dot\chi^4 \Ub' + \Ueff' \ .
\eea
If $\a (1+8\a \Ub)\dot\chi^2\ll1$, these reduce to the usual FRW equations for a minimally coupled field with a canonical kinetic term and potential $\Ueff$.\footnote{The results below show that in slow-roll
$\a (1+8\a \Ub)\dot\chi^2=\frac{2}{3} \a \Ub \epsilon = \frac{2}{3} \frac{\a \Ub}{1+8\a \Ub} \epsb< \frac{1}{12} \epsb$, so this condition is satisfied.} Slow-roll inflation is then described by the slow-roll parameters, the first of which are
\bea \label{SR}
  \epsilon &=& \ha \left( \frac{\Ueff'}{\Ueff} \right)^2 =  \frac{1}{1+8\a \Ub} \epsb \el
   \eta &=& \frac{\Ueff''}{\Ueff} = \etab - 3 \frac{8 \a \Ub}{1+8\a \Ub} \epsb \el
   \zeta &=& \frac{\Ueff'}{\Ueff} \frac{\Ueff'''}{\Ueff} = \zetab + 12 \frac{(8 \a \Ub)^2}{(1+8\a \Ub)^2} \epsb^2 - 8 \frac{8 \a \Ub}{1+8\a \Ub} \epsb \etab \ .
\eea
The effect of the non-minimal coupling $G$ on inflation in the Palatini case is well-established \cite{Bauer:2008zj, Bauer:2010jg, Rasanen:2017, Enckell:2018, Markkanen:2017, Jarv:2017, Tenkanen:2017, Carrilho:2018}. We have therefore focused on the new terms involving $8\a \Ub$, and written the slow-roll parameters relative to what they are in the case $\a=0$ (denoted by $\epsb$, $\etab$, and $\zetab$). For the amplitude of perturbations, scalar spectral index, tensor-to-scalar ratio, and running of the scalar spectral index we obtain
\bea
  24 \pi^2 A_s &=& \frac{\Ueff}{\epsilon} = \frac{\Ub}{\epsb} \el
  n_s-1 &=& 2 \eta - 6 \epsilon = 2 \etab - 6 \epsb \el
  r &=& 16 \epsilon = \frac{16}{1+8\a \Ub} \epsb \el
  \a_s &=& 16 \epsb \etab - 24 \epsb^2 - 2 \zetab \ .
\eea
Although the $R^2$ term modifies $\epsilon$, $\eta$ and $\zeta$, the effect cancels out in $A_s$, $n_s$ and $\a_s$ to leading order in the slow-roll parameters. (The number of e-folds is also unaffected.) Observables defined as higher order  derivatives of the power spectrum (such as the running of the running) are also unaffected to leading order, because the slow-roll scalar power spectrum remains the same:
\bea
  \mathcal{P}_\mathcal{R}=\frac{\Ueff}{24\pi^2\epsilon}=\frac{\Ub}{24 \pi^2 \epsb} \ .
\eea
When $1+8\a \Ub$ grows the change in the effective potential $\Ueff$ and its slope is compensated by the change in the field $\chi$ given by \eqref{ft}. However, that is not the case for the tensor power spectrum
\bea
 \mathcal{P}_T = \frac{2 \Ueff}{3 \pi^2} = \frac{2\Ub}{3 \pi^2 (1 + 8 \a \Ub)} \ .
\eea
Thus we have $r \rightarrow r/(1+8\a \Ub)$, and the tensor-to-scalar ratio can be decreased without limit by increasing $\a$. The tensor spectral index is also suppressed by the same factor, $n_T \rightarrow n_T/(1+8\a \Ub)$.

\section{Conclusions} \label{sec:conc}

We have studied inflation with a scalar field in the Palatini formulation when the action contains an $R^2$ term. The introduction of the $R^2$ term modifies the relation between the curvature and the scalar field: it makes the effective potential \eqref{Ueff} asymptotically flat and lowers its value. The change of the slope and the absolute value of the potential is (to leading order in the slow-roll parameters) balanced by a modification of the rate of change of the scalar field in the scalar spectral index and the scalar amplitude of perturbations. However, the decrease in the absolute value translates into a suppression of the tensor-scalar-ratio $r$ and the tensor spectral index $n_T$ by the factor $1/(1+8\a \Ub)$, where $\Ub$ is the inflaton potential in the Einstein frame when $\a=0$. A more complicated action, such as a more involved form of $F(R)$ or a different coupling between the scalar field and gravity, could lead to more drastic changes, as studied in \cite{Kijowski:2016, Afonso:2017, Afonso:2018a, Afonso:2018b, Afonso:2018c}.

The difference between the metric formulation and the Palatini formulation in inflation has been previously studied in the presence of a non-minimal coupling of the inflaton to gravity \cite{Bauer:2008zj, Bauer:2010jg, Rasanen:2017, Enckell:2018, Markkanen:2017, Jarv:2017, Tenkanen:2017, Carrilho:2018}, especially in the context of Higgs inflation, where the main effect is to decrease $r$ by several orders of magnitude. We have shown that an $R^2$ term has the same effect. This is in contrast to the metric case, where an $R^2$ term introduces a new scalar degree of freedom. (The new degree of freedom has been argued to destabilise Higgs inflation and require fine-tuning in the metric formulation, except in critical Higgs inflation \cite{Salvio:2015kka, Salvio:2017oyf}; in the Palatini formulation there is no such problem.)

It has been pointed out in the context of Higgs inflation that the specification of the gravitational degrees of freedom (in the case of Palatini formulation, whether the connection is an independent variable) is an extra choice for models of inflation, which has to be made before they can be compared to observations \cite{Rasanen:2017, Enckell:2018}. We see that this is not only an issue for models with a non-minimal coupling between the inflaton and gravity, but also models with an extended gravitational action.

The $R^2$ term should generally be included in the action, and the mechanism we have discussed can be used to lower $r$ for any scalar field model of inflation. It can bring a model that predicts $r$ in excess of the observational upper bound but otherwise fits the data, such as the simplest chaotic inflation model with the potential $V(h)=\frac{1}{2}m^2 h^2$, back into agreement with observations.

\acknowledgments

We thank Tommi Tenkanen for participating in the beginning of this project. LPW is supported by Magnus Ehrnrooth Foundation.

\bibliographystyle{JHEP}
\bibliography{star}


\end{document}